\documentclass[prl,aps,superscriptaddress,footinbib,twocolumn]{revtex4-2}
\usepackage[T1]{fontenc}
\usepackage{courier}
\usepackage{url}
\usepackage{xcolor}
\usepackage{amsmath}
\usepackage{amssymb}
\usepackage{stmaryrd}
\usepackage{graphicx}
\usepackage{bm}
\usepackage{dcolumn}
\usepackage{multirow}
\usepackage{esint}
\usepackage{ulem}
\usepackage[ruled,vlined]{algorithm2e}
\usepackage{algpseudocode}
\usepackage{float}
\usepackage{lineno}

\DeclareMathOperator{\Tr}{Tr}
\usepackage[unicode=true,bookmarks=true,bookmarksnumbered=false,bookmarksopen=false,breaklinks=false,pdfborder={0 0 1},backref=false,colorlinks=true]
 {hyperref}
\hypersetup{
 linkcolor=magenta, urlcolor=blue, citecolor=blue, pdfstartview={FitH}, hyperfootnotes=false, unicode=true}

\usepackage{todonotes}

\begin{document}

\title{Quantum ensemble learning with a programmable superconducting processor}
\affiliation{School of Physics, ZJU-Hangzhou Global Scientific and Technological Innovation Center, and Zhejiang Key Laboratory of Micro-nano Quantum Chips and Quantum Control, Zhejiang University, Hangzhou 310027, China\\
$^{2}$ Institute of Fundamental and Frontier Sciences, University of Electronic Science and Technology of China, Chengdu, 610051, China\\
$^{3}$ School of Physics and Key Laboratory of Quantum State Construction and Manipulation (Ministry of Education), Renmin University of China, Beijing 100872, China\\
$^{4}$ Graduate School of China Academy of Engineering Physics, Beijing 100193, China
}

\author{Jiachen Chen$^{1}$}\thanks{These authors contributed equally to this work.}
\author{Yaozu Wu$^{1}$}\thanks{These authors contributed equally to this work.}
\author{Zhen Yang$^{2}$}\thanks{These authors contributed equally to this work.}
\author{Shibo Xu$^{1}$}
\author{Xuan Ye$^{2}$}
\author{Daili Li$^{3}$}
\author{Ke Wang$^{1}$}
\author{Chuanyu Zhang$^{1}$}
\author{Feitong Jin$^{1}$}	
\author{Xuhao Zhu$^{1}$}
\author{Yu Gao$^{1}$}
\author{Ziqi Tan$^{1}$}
\author{Zhengyi Cui$^{1}$}
\author{Aosai Zhang$^{1}$}	
\author{Ning Wang$^{1}$}
\author{Yiren Zou$^{1}$}
\author{Tingting Li$^{1}$}
\author{Fanhao Shen$^{1}$}
\author{Jiarun Zhong$^{1}$}
\author{Zehang Bao$^{1}$}
\author{Zitian Zhu$^{1}$}
\author{Zixuan Song$^{1}$}
\author{Jinfeng Deng$^{1}$}
\author{Hang Dong$^{1}$}
\author{Pengfei Zhang$^{1}$}
\author{Wei Zhang$^{3}$}
\author{Hekang Li$^{1}$}
\author{Qiujiang Guo$^{1}$}
\author{Zhen Wang$^{1}$}
\author{Ying Li$^{4}$}\email{yli@gscaep.ac.cn}
\author{Xiaoting Wang$^{2}$}\email{xiaoting@uestc.edu.cn}
\author{Chao Song$^{1}$}\email{chaosong@zju.edu.cn}
\author{H. Wang$^{1}$}

\begin{abstract}

Quantum machine learning is among the most exciting potential applications of quantum computing.
However, the vulnerability of quantum information to environmental noises and the consequent high cost for realizing fault tolerance has impeded the quantum models from learning complex datasets.
Here, we introduce Adaboost.Q, a quantum adaptation of the classical adaptive boosting (Adaboost) algorithm designed to enhance learning capabilities of quantum classifiers. 
Based on the probabilistic nature of quantum measurement, the algorithm improves the prediction accuracy by refining the attention mechanism during the adaptive training and combination of quantum classifiers.
{We experimentally demonstrate the versatility of our approach on a programmable superconducting processor, where we observe notable performance enhancements across various quantum machine learning models, including quantum neural networks and quantum convolutional neural networks.}
With Adaboost.Q, we achieve an accuracy above $86\%$ for a ten-class classification task over 10,000 test samples, and an accuracy of $100\%$ for a quantum feature recognition task over 1,564 test samples.
Our results demonstrate a foundational tool for advancing quantum machine learning towards practical applications, which has broad applicability to both the current noisy and the future fault-tolerant quantum devices.
\end{abstract}

\maketitle

The intersection between machine learning and quantum computing gives rise to the field of quantum machine learning that has attracted considerable attention.
With the exponentially large Hilbert space, a quantum computer promises to offer representational power for recognizing complex data patterns that are challenging to recognize classically~\cite{Biamonte2017_quantum, Dunjko2018Machine, DasSarma2019Machine, Cerezo2022Challenges, Havlicek2019_supervised, Liu2021_aRigorous}.
Recently, the power of quantum learning models has been extensively studied in terms of quantum neural network (QNN) and quantum convolutional neural network (QCNN)~\cite{Huang2021Power, Jerbi2023Quantum, Cong2019Quantum} and, with the rapid advances of quantum technologies across various physical platforms~\cite{Arute2019Quantum, Kim2023_Evidence, Xu2023nonAbelian, Bluvstein2024_logical, Iqbal2024_nonAbelian}, supported by a growing number of experiments~\cite{Huang2022_quantum, Herrmann2022_realizing, GONG2023_quantum, kornjaca2024_largescale, zhang2024_quantumcontinual}.
Yet, the sizes of datasets used in these demonstrations remain small compared with those typically used in classical machine learning.

As a crucial step towards practical applications, it is desirable to test the quantum learning models on datasets with large sizes.  
However, when faced with large-scale datasets, the task becomes challenging, necessitating the development of additional methods to boost the performance of learning models. 
By leveraging the power of joint decision, ensemble methods have achieved dramatic success in improving the accuracy of classical machine learning models in the last three decades~\cite{Freund1995_aDecision, Schapire1999_improved, Hastie2009_multi, Sagi2018_ensemble, Viola2004_robust}.
{More recently, theoretical work suggests that ensemble methods can improve the predictive performance of QNNs~\cite{Li2024_ensemble}.}

In this work, we present a quantum version of the adaptive boosting (AdaBoost) algorithm, dubbed AdaBoost.Q.
{On a superconducting quantum processor, we demonstrate using the algorithm to improve the performance of quantum models in two supervised learning experiments.}
The first experiment realizes a ten-class classification of the MNIST handwritten digits dataset~\cite{Deng2012_theMNIST} with a ten-qubit QNN classifier.
By employing Adaboost.Q, we improve the testing accuracy from $80\%$ to above $86\%$ over the full-size MNIST test dataset.
The second experiment aims to classify three quantum phases of a spin chain model with a $15$-qubit QCNN classifier.
We show that the performance of the QCNN classifier can be significantly improved by Adaboost.Q, with the testing accuracy enhanced from $77\%$ to $100\%$ over 1,564 test samples.
Our results provide a widely applicable method for pushing quantum machine learning towards practical applications.

\begin{figure}[t]
\includegraphics[width=0.5\textwidth]{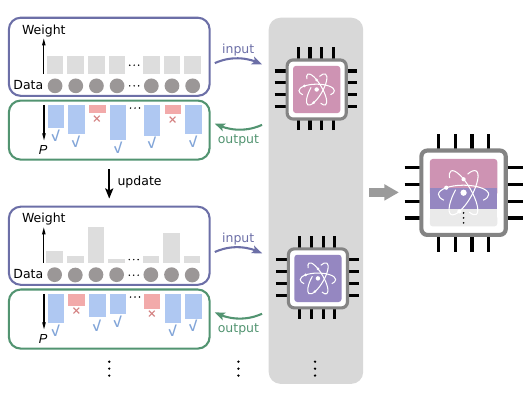}
\caption{
\label {Fig1}
{\bf Schematic diagram of AdaBoost.Q}. The algorithm is designed to generate a strong quantum classifier (right) by combining multiple weak quantum classifiers (middle).
Each weak quantum classifier takes the same training data as an input, and it outputs the classification result of each data sample along with a probability $P$, which characterizes the confidence of the prediction.
The weak quantum classifiers are trained iteratively, using reweighted versions of the training set, with the weights depending on the correctness of the predictions and finely tuned by the output probabilities of the previous classifier.
This allows the subsequent quantum classifiers to focus on samples that were not well classified previously.
The sample weights for the first classifier are assigned evenly among the training set.
}
\end{figure}
\vspace{.5cm}
\noindent\textbf{\large{}Framework and experimental setup}

It is a common human practice to aggregate and weigh different opinions to make a complex decision.
The ensemble methodology extends this idea to the world of machine learning, aiming to construct a highly accurate classifier (referred to as a ``strong'' classifier) by combining multiple ``weak'' classifiers, each of which may only slightly outperform a random guess.
The Adaboost algorithm is among the most prominent ensemble methods to generate a strong classifier~\cite{Freund1995_aDecision, Schapire1999_improved, Sagi2018_ensemble}.
It works by training the weak classifiers sequentially on the same training set.
At the core of the Adaboost algorithm is an attention mechanism, where more attention is paid to data points that were previously misclassified when training the subsequent classifier.
The level of attention paid is determined by a sample weight that is assigned to each training point.
After the training procedure, each weak classifier is also assigned a weight for scoring its importance in forming the strong classifier.
In the Adaboost.Q algorithm, we extract these weights based on the probabilistic nature of the quantum classifiers.

We consider quantum classifiers that are built with parameterized quantum circuits.
For a supervised $K$-class classification task, the training set consists of pairs of datasets, $\{{x}_i, y_i\}_{i=1}^{N}$, where ${x}_i$ represents the data sample, $y_i$ indexes the corresponding class label, and $N$ is the training size.
To classify the data samples, we select $m$ qubits, with $m\geq \lceil \log_{2}K \rceil$, from the quantum classifier and measure them in the computational basis, which can be described by a set of basis projectors $\{\Pi_j\}_{j=0}^{2^m-1}$ known as the projection-valued measures (PVM).
We divide the PVM equally into $K$ groups, with the $k$th group containing projectors indexed from $k\lfloor 2^m/K\rfloor$ to $(k+1)\lfloor 2^m/K\rfloor-1$, while discarding the last $2^m-K\lfloor 2^m/K\rfloor$ projectors.
The data sample is classified as the $k$th class if the measured state is located in the $k$th group.

According to Born's rule, for an input sample $x_i$, the probability of measuring the basis state in the $k$th group is $P_{k}(x_i, \bm{\theta})=\sum_{j=k\lfloor 2^m/K\rfloor}^{(k+1)\lfloor 2^m/K\rfloor-1}\Tr(\rho(x_i, \bm{\theta})\Pi_j)$, where $\rho(x_i, \bm{\theta})$ denotes the reduced density matrix of the $m$ measured qubits of the quantum classifier parameterized by $\bm{\theta}$.
The corresponding predicted label of the quantum classifier is obtained as $\Tilde{y}_i = \mathop{\arg\max}\limits_{k} P_{k}(x_i, \bm{\theta})$.
Note that the probability $P(x_i, \bm{\theta}) = \mathop{\max}\limits_{k} P_{k}(x_i, \bm{\theta})$ naturally characterizes the confidence of the prediction.

\begin{figure*}[!t]
\centering
\includegraphics[width=1.0\textwidth]{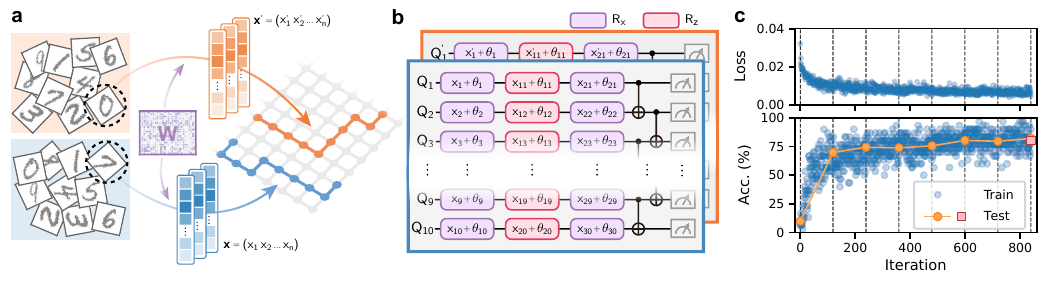}
\caption{
\label {Fig2}
{\bf Ten-class classification of the MNIST dataset with QNN.}
\textbf{a}, 
Experimental setup.
{The training set is composed of 3,600 MNIST handwritten digit images, each with a size of $28 \times 28$ pixels.
An image is transformed into a 30-dimensional vector $\bm x$ using a trainable matrix $\bm W$ for further quantum encoding.}
At each training step, we select a batch of 30 images to train the classifier, which is split evenly into two {sub-batches}, denoted here by $\bm x$ and $\bm x'$, and fed to {two copies of the QNN} in parallel.
Each {copy} is constructed with a ten-qubit chain selected from the quantum processor.
\textbf{b}, The two QNN circuits, each of which is composed of three layers of single-qubit gates followed by two layers of CNOT gates. The rotation angles of single-qubit gates are used to encode the data vector and trainable parameters $\bm{\theta}$. Here $R_x$ and $R_z$ denote the single-qubit rotation gates around the x- and z-axis, respectively.
\textbf{c}, Loss function (top) and accuracy for the test and training sets (bottom) at each training step. 
The training is carried out for 7 epochs, with each epoch consisting of 120 training steps as separated by the dash lines.
After each epoch, the QNN is monitored with 1,000 images randomly selected from the MNIST test set (orange circle dots).
At the end of the training, we measure the test accuracy over the whole MNIST test set containing 10,000 images (red square dot).
}
\end{figure*}

Using the probability information, we establish a refined criterion for calculating the weights of the data samples and classifiers following the spirit of the real Adaboost algorithm~\cite{Schapire1999_improved, Schapire2003_theBoosting}.
Specifically, we initialize all the sample weights to be $1/N$ when training the first classifier.
During the iterative training of the subsequent classifiers, the sample weights are updated according to
\begin{equation}
    w_{l+1, i} = \frac{w_{l, i}}{Z_{l+1}}\text{exp}[P(x_i, \bm{\theta}_l^*) \cdot (1 - 2 \delta_{\Tilde{y}_{l, i}y_i})], \label{Adaboost_Sample_Weight}
\end{equation}
where $i$ indexes the samples, $l$ indexes the weak classifiers, $\bm{\theta}_l^*$ is the optimal parameter for the $l$th classifier, $\delta$ denotes the Kronecker delta, and $Z_{l+1}=\sum_{i=1}^{N}w_{l, i}\,\text{exp}[P(x_i, \bm{\theta}_l^*) \cdot (1 - 2 \delta_{\Tilde{y}_{l, i}y_i})]$ is the normalizing factor.
The loss function of the $l$th classifier is given by $\mathcal{L}_l = \sum\limits_{b=1}^{B}\mathcal{L}_{l, b}$, with $B$ denoting the batch size and 
\begin{equation}
\label{qnn_loss}
{\mathcal{L}_{l, b}=-w_{l, b}\,\text{ln}\,[ P_{y_b}({x_b}, \bm{\theta})]}.
\end{equation}

After training the $l$th classifier, we calculate its weight as
\begin{equation}
    \alpha_l = \ln{\frac{c_l\mathcal{P}_l^{\text{true}}}{\mathcal{P}_l^{\text{false}}}}, \label{Adaboost_Classifier_Weight}
\end{equation}
where $\mathcal{P}_l^{\text{true}} = \sum_i w_{l,i}P(x_i, \bm{\theta}_l^*)\delta_{\Tilde{y}_{l,i}y_i}$ and $\mathcal{P}_l^{\text{false}} = \sum_i w_{l,i}P(x_i, \bm{\theta}_l^*)(1 - \delta_{\Tilde{y}_{l,i}y_i})$.
The additional parameter $c_l$, which takes the value of unity by default, can be slightly tuned to optimize the training accuracy in practice. 
The ensemble classifier is constructed by combining all the trained weak classifiers (also referred to as base classifiers), which classifies $x_i$ according to
\begin{equation}
    \Tilde{y}_i = \mathop{\arg\max}\limits_{k} \sum_l \alpha_lP_{k}(x_i, \bm{\theta}_l^*).
\end{equation}
{The workflow of Adaboost.Q is illustrated in Fig.~\ref{Fig1}. }

We experimentally demonstrate the effectiveness of our approach on a fully programmable superconducting quantum processor~\cite{Xu2023nonAbelian}.
The qubits on the processor are of the frequency-tunable transmon type, which are arranged in an $11\times11$ square lattice, with the neighboring qubits connected by tunable couplers.
{For the ten-class classification of the MNIST dataset, we select 20 qubits to construct two copies of a ten-qubit QNN classifier, which run in parallel to accelerate training}. 
For the quantum data classification task, we build a QCNN classifier with a carefully designed one-dimensional (1D) chain consisting of 15 qubits.
{All the classifiers are essentially variational quantum circuits compiled into the native gate sets, i.e., the parameterized single-qubit gates and two-qubit CZ gates between neighboring qubits.}
The median Pauli error rates of the parallel single- and two-qubit gates, characterized with the simultaneous cross-entropy benchmarking technique, are around $5\times10^{-4}$ and $6\times10^{-3}$, respectively.
See Supplementary Sec.~IIA for details on device and gate performances.

\vspace{.5cm}
\noindent\textbf{\large{}Ensemble learning with QNN}

As a first demonstration, we apply our method to improve the performance of QNN-based classifiers, which have been intensively studied in recent years~\cite{schuld2020_circuit, Ren2022Experimental, Li2022_quantum, Jerbi2023Quantum}.
We focus on a ten-class classification task with the MNIST handwritten digit dataset, which is widely used in benchmarking machine learning models.
This task had been challenging for quantum hardware, and it was not until recently that an experimental test accuracy of around $62\%$ over 500 test samples was reported~\cite{kornjaca2024_largescale}.

We use a chain of ten qubits to construct the QNN circuit with 30 trainable parameters.
The training dataset contains 3,600 $28\times28$-pixel {images} (360 for each digit) selected from the MNIST dataset.
To encode the classical data, we adapt the encoding scheme from the end-to-end learning framework~\cite{pan2023_experimental}. 
Specifically, we first vectorize the image data and then transform it to an array of rotational angles $\bm{x}$ with a transform matrix $\bm{W}$, following which we encode them into the single-qubit rotation of the QNN circuit alternatively with the trainable parameters $\bm{\theta}$ (Fig.~\ref{Fig2}a).
{Both $\bm{W}$ and $\bm{\theta}$ are trained simultaneously to minimize the loss function during the learning procedure.}
{To reduce the runtime, we further parallelize quantum computing by constructing and running two copies of the QNN simultaneously on the processor.}

The training procedure of a single QNN classifier is exemplified in Fig.~\ref{Fig2}c, where the loss function converges to about 0.007 after 840 training steps.
At the end of the training procedure, we input all 10,000 samples of the MNIST test dataset to the trained classifier, 
obtaining an overall testing accuracy around $80.5\%$, which is consistent with the training accuracy, verifying the generalizability of the trained QNN model. See Methods for the details about the training procedure.

\begin{figure}[h]
\includegraphics[width=0.5\textwidth]{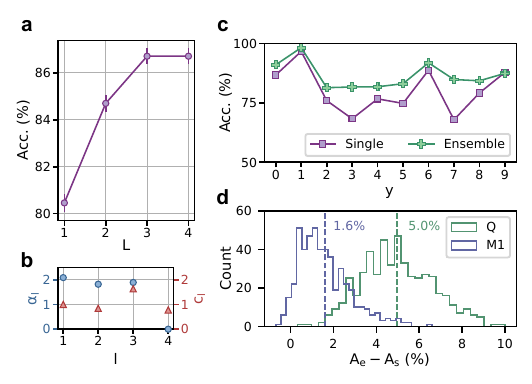}
\caption{
\label {Fig3}
{\bf Experimental implementation of Adaboost.Q.}
\textbf{a}, Testing accuracy of the ensemble classifiers composed of $L$ base classifiers, with $L=1$ to $4$. The error bars are obtained as the standard deviations from bootstrapping (resampled 1,000 times from the original data sets). The accuracy is estimated based on the whole MNIST test set.
\textbf{b}, The experimental weight $\alpha_l$ and adjustment parameter $c_l$ of each base classifier $l$.
\textbf{c}, Comparison of the classification performance on the whole MNIST test set with single and ensemble classifiers, where the accuracy of classifying each digit is displayed.
\textbf{d}, Comparison of the performance between the conventional Adaboost.M1 and Adaboost.Q with numerical simulation considering the same task, where both ensemble classifiers are composed of four base classifiers.
The simulation is performed 500 times each. The histogram of the accuracy improvement, $A_e-A_s$ with $A_e$ and $A_s$ being the testing accuracy of the ensemble and single classifiers, is plotted, with the average values denoted by the dash lines.
}
\end{figure}

With the performance of the base QNN classifier established, {we move on to Adaboost.Q}.
In Fig.~\ref{Fig3}a, we plot the testing accuracies of the ensemble QNN classifiers during the implementation of Adaboost.Q.
A notable increase of the accuracy is observed, from $80.5\%$ to {$86.7\%$} over the first two iterations, after which the accuracy saturates.
The observation is consistent with the weight calculated for each base classifier, which dramatically drops to near zero for the fourth base classifier (Fig.~\ref{Fig3}b).
In Fig.~\ref{Fig3}c, we compare the classification results with and without using Adaboost.Q, observing an improvement of testing accuracy for the ensemble classifier across almost all digits.
We also perform numerical simulation to compare our method with the conventional Adaboost.M1 algorithm~\cite{FREUND1997_aDecision} and observe that the former shows better performance, as shown in Fig.~\ref{Fig3}d. See Supplementary Sec.~I for details of the numerical simulation and the Adaboost.M1 algorithm.

\begin{figure*}[t]
\centering
\includegraphics[width=1.0\textwidth]{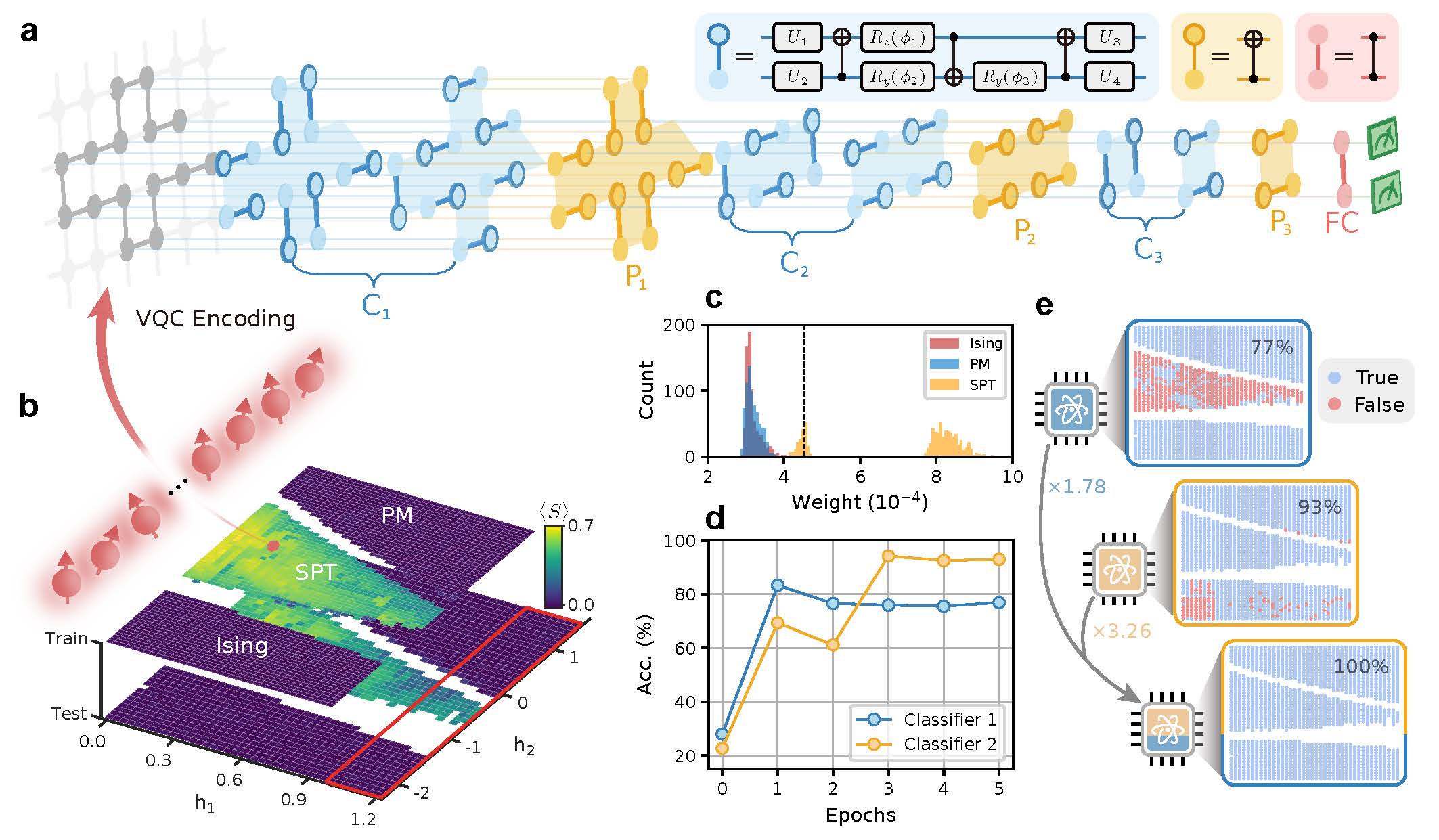}
\caption{
\label {Fig4}
{\bf Ensemble learning of quantum data with QCNN.}
\textbf{a}, Structure of the QCNN circuit, which consists of three alternating convolutional ($\textsf{C}_{\textsf{i}}$) and pooling ($\textsf{P}_{\textsf{i}}$) layers, followed by a fully-connected ($\textsf{FC}$) layer.
Each convolutional layer contains two layers of convolutional kernels, each of which consists of seven single-qubit gates with 15 variational parameters and three two-qubit CNOT gates.
We apply the convolutional kernels in a translationally invariant way, resulting in a total of 45 variational parameters.
The pooling layer applies a layer of two-qubit CNOT gates and then passes the target qubits to the {subsequent} layer, thus reducing the qubit number by half. 
After three rounds of pooling, we are left with two qubits, on which we apply a CZ gate as a fully connected layer. The two qubits are then measured in the computational basis for the classification task.
\textbf{b}, The experimentally measured string order parameters $\langle S \rangle$ for the quantum states in the training and test sets, respectively.
Each quantum state is {the ground state} of a cluster-Ising Hamiltonian with 15 spins, which {has} two parameters $h_1$ and $h_2$ (see Methods).
The states can be {approximately} prepared with variational quantum circuits and directly input into the QCNN classifier.
The test set is generated in a parameter regime larger than that for the training set, {and we highlight their difference with the red box}.
\textbf{c}, Histogram of the sample weights for training the second QCNN base classifiers.
The weights for the quantum states in different phases are plotted separately.
The dashed line represents the initial weight that is assigned to all data samples when training the first base classifier.
\textbf{d}, Testing accuracy of the first (blue) and the second (yellow) base classifier at each epoch.
\textbf{e}, Classification results of the first (top), second (middle), and ensemble (bottom) classifiers on the test set. {The classifier weights are $1.78$ and $3.26$, respectively.}
} 
\end{figure*}

\vspace{.5cm}
\noindent\textbf{\large{}Ensemble learning with QCNN}

To demonstrate the versatility of Adaboost.Q, we further apply it to QCNN-based classifiers.
The task is to classify the ground states of a cluster-Ising Hamiltonian~\cite{SPT_model_ref1}, which can either belong to a symmetry-protected topological (SPT) phase, a paramagnetic (PM) phase, or an Ising phase (Methods).
Previous studies~\cite{Cong2019Quantum, Herrmann2022_realizing} have revealed the advantages of QCNN over the direct measurement of the string order parameter
in identifying the SPT phase.
Here, we consider a more complex task, i.e., the classification of all three phases.
{The QCNN circuit is implemented on a 1D chain of $15$ qubits on our device. QCNN can achieve an exponential reduction of trainable parameters compared with the generic QNN circuit \cite{Cong2019Quantum}, at the cost of involving long-range two-qubit gates along the chain.} We circumvent this challenge by carefully designing the topology of the selected 1D chain, such that all {necessary} two-qubit gates can be directly implemented, as shown in Fig.~\ref{Fig4}a.

A QCNN classifier is typically composed of convolutional, pooling, and fully-connected layers.
In our experiment, we construct the convolutional layer with two layers of convolutional kernels applied in a translationally invariant way.
Each convolutional kernel is a parameterized two-qubit $U(4)$ unitary~\cite{U4_gate_ref1}.
A pooling layer applies parallel controlled-X (CNOT) gates and then discards the control qubits.
After performing the convolutional and pooling layers for three rounds, we implement a CZ gate on the remaining two qubits as the fully connected layer.
The two qubits are then readout to obtain the classification results.

We use variational quantum circuits to prepare approximate ground states for constructing the training and test datasets.
The quantum data can be directly input into the QCNN circuit as initial states.
Within the reach of variational quantum circuits, our training (test) dataset consists of 2,204 (1,564) points spanning the phase diagram of the cluster-Ising model, with the experimentally measured string order parameters shown in Fig.~\ref{Fig4}b. To examine the generalization of the QCNN classifier, we extend to the parameter regime in generating the test samples (red box in Fig.~\ref{Fig4}b).
See Methods and Supplementary Sec.~IIB for more details about the generation of training and test sets.

In Fig.~\ref{Fig4}c-e, we show the performance of Adaboost.Q. 
For a single QCNN classifier, we find that the testing accuracy for the three-class classification task is limited to around $80\%$.
A detailed analysis of the classification results reveals that the states in the SPT phase are most likely to be misclassified (Fig.~\ref{Fig4}e, top panel).
By tuning the weights of the training states (Fig.~\ref{Fig4}c), the second classifier achieves a remarkable increase in the testing accuracy, reaching a value of $93\%$, with most of the misclassified samples located in the {Ising} phase (Fig.~\ref{Fig4}e, middle panel).
Finally, by combining only two base classifiers, the ensemble classifier achieves a testing accuracy of $100\%$ on the whole test set (Fig.~\ref{Fig4}e, bottom panel), demonstrating the high efficiency of our algorithm.

\vspace{.5cm}
\noindent\textbf{\large{}Conclusion}

\noindent 

In this work, we propose Adaboost.Q, a quantum ensemble learning algorithm that refines the data and classifier weights based on the probabilistic nature of quantum outputs.
The protocol outperforms the conventional Adaboost.M1 algorithm and is examined experimentally with multi-class classification tasks on both classical and quantum data.
In particular, we achieve the best-known testing accuracy for the ten-class classification of the MNIST handwritten digits dataset on quantum hardware, obtained by testing the classifier over the full MNIST test set.
Our approach is generally applicable to different quantum learning models and across different physical platforms in both the current noisy intermediate-scale quantum era and the coming fault-tolerant quantum era, which we anticipate would benefit the future exploration of quantum learning advantages.

\vspace{.6cm}
\noindent\textbf{\large{}Data availability}\\
The data presented in the figures and that support the other findings of this study will be made publicly available for download on Zenodo/Figshare/Github upon publication.

\vspace{.6cm}
\noindent\textbf{\large{}Code availability}\\
The numerical simulation codes for this study will be made publicly available for download on Github upon publication.

\vspace{.5cm}
\noindent\textbf{Acknowledgement} 
{The device was fabricated at the Micro-Nano Fabrication Center of Zhejiang University.  We acknowledge support from the National Key R\&D Program of China (Grant No.~2023YFB4502600), 
the National Natural Science Foundation of China (Grant Nos.~12174342, 92365301, 12274367, 12322414, 12274368, 12225507, 12088101, 12074428 and 92265208), and the Zhejiang Provincial Natural Science Foundation of China (Grant Nos.~LDQ23A040001, LR24A040002). Y.L. is also supported by NSAF (Grant No. U1930403).
}

\vspace{.3cm}
\noindent\textbf{Author contributions}  
X.W. and Y.L. initiated the study. J.C. and Y.W. carried out the experiments under the supervision of C.S.. J.C. and X.Z. designed the device and H.L. fabricated the device, supervised by H.W.. J.C., Z.Y., Y.L., and D.L. performed the numerical simulation, supervised by W.Z., X.W., and C.S.. J.C., Z.Y., X.Y., D.L., Y.L., and X.W. conducted the theoretical analysis. C.S., Y.W., J.C., and Y.L. co-wrote the manuscript. J.C., Y.W., S.X., K.W., C.Z., F.J., X.Z., Y.G., Z.T., Z.C., A.Z., N.W., Y.Z., T.L., F.S., J.Z., Z.B., Z.Z., Z.S., J.D., H.D., P.Z., H.L., Q.G., Z.W., C.S., and H.W. contributed to the experimental setup. All authors contributed to the analysis of data, the discussions of the results, and the writing of the manuscript.

\vspace{.5cm}
\noindent\textbf{\large{Methods}}
\newline\noindent
\noindent\textbf{Data generation}.
\newline\noindent
The training and test sets used for the ten-class classification task originate from the MNIST handwritten digits dataset.
To accelerate the training procedure, we use the $k$-means algorithm~\cite{kmeans} to select 3,600 representative images from the MNIST training dataset to form the training set.
Specifically, we use KMeans function of the scikit-learn package~\cite{sklearn_api} to divide each number into 360 clusters. 
We select one image from each cluster to form ten numbers totalling 3,600 images. 
During the clustering process of KMeans, we perform 180 random initial centroids and select the case the clustering with the best inertia.
The test set is constructed with all 10,000 images from the MNIST test dataset.
{During the training procedure, we also randomly select 1,000 images, with 100 for each digit, from the test set to monitor the testing accuracy.}

For the quantum phase recognition task, the quantum dataset consists of ground states of the cluster-Ising Hamiltonian:
\begin{equation}
    H = -\sum_{i=1}^{N-2}Z_{i}X_{i+1}Z_{i+2} - h_{1}\sum_{i=1}^{N}X_{i} - h_{2}\sum_{i=1}^{N-1}X_{i}X_{i+1}, \label{spt_hamiltonian}
\end{equation}
where $N=15$ in our system. $\left\{X_{i}, Z_{i}\right\}$ are the Pauli operators acting on the $i$th spin.
Depending on the choice of the two model parameters $\{h_1,h_2\}$, the ground states can belong to either the SPT phase, the PM phase, or the Ising phase.
The SPT phase can be distinguished by measuring the string order parameter:
\begin{equation}
    S = Z_{1}X_{4}...X_{12}X_{14}Z_{15},
    \label{SOP}
\end{equation}
as shown in Fig.~\ref{Fig4}b.
To construct the training (test) set, we select $2,204$ ($1,564$) ground states in the parameter regimes of $h_1\in[0,1)$ ($h_1\in[0,1.2]$) and {$h_2\in(-2.3,1.6]$}, respectively.
For a given $\{h_1,h_2\}$, we use a variational quantum circuit (VQC) to prepare the corresponding ground state.
The circuit is trained on a classical computer before being deployed on the quantum processor {(See Supplementary Sec.~IIB for details on the training procedure)}. 

\vspace{.3cm}
\noindent\textbf{Algorithms for quantum ensemble learning}

\noindent The AdaBoost.Q algorithm proposed in this work is shown in Algorithm~\ref{alg:AdaBoost}.

\begin{algorithm}
	\caption{AdaBoost.Q}
	\label{alg:AdaBoost}
	\SetAlgoLined
	\KwIn{~\\
		Training set $\mathcal{T}=\left\{(x_i, y_i)\right\}^{N}_{i=1}$, where $x_i \in \mathcal{X}$ and $y_i \in \mathcal{Y} \subseteq \left\{1, 2, ..., K\right\}$;\\
		The number of weak classifier $L$; \\
		A series of weak classifiers $\left\{G(x, \bm{\theta}_l)\right\}^{L}_{l=1}$,
		\begin{equation}
			G(x, \bm{\theta}_l): \mathcal{X} \rightarrow \bm{R}^K, \quad G(x_i, \bm{\theta}_l) = \left\{P_{k}(x_{i}, \bm{\theta}_l)\right\}_{k=1}^{K} \nonumber,
		\end{equation}
		with $P_{k}(x_i, \bm{\theta}_l)$ defined in the main text.
	}
	\KwOut{~\\
		Combined classifier
		\begin{equation}
			\mathcal{G}(x) = \sum_{l=1}^{L}\alpha_l \cdot G(x, \bm{\theta}^*_l) \nonumber,
		\end{equation}
		where $\alpha_{l}$ and $\bm{\theta}^*_l$ are the weight and optimal parameters of the $l$th classifier.
		The output label of the combined classifier is
		\begin{equation}
			\tilde{y}_i = \mathop{\arg\max}_{k}\mathcal{G}(x_i) \nonumber.
		\end{equation} 
	}
	\hrulefill ~\\
	
	Initialize the weights of the training set,
	\begin{equation}
		\bm{w}_1 = \left\{w_{1,i}\right\}_{i=1}^{N}, \nonumber
	\end{equation}
	where $w_{1, i} = 1 / N$.
	
	\For{$l = 1:L$}{
		Train the classifier $G(x, \bm{\theta})$ with the weighted training set.
		Obtain
		\begin{equation}
			\tilde{y}_{l, i} = \mathop{\arg\max}\limits_{k}G(x_i, \bm{\theta}^{*}_{l}) \nonumber,
		\end{equation}
		with probability:
		\begin{equation}
			P(x_i, \bm{\theta}^{*}_{l}) = \mathop{\max}G(x_i, \bm{\theta}^{*}_{l}) \nonumber.
		\end{equation}
		
		Calculate the $l$th classifier's weight:
		\begin{equation}
			\alpha_l = \text{ln}\frac{c_l\mathcal{P}^{\text{true}}_{l}}{\mathcal{P}^{\text{false}}_{l}} \nonumber,
		\end{equation}
		with
		$$
		c_l = 
		\begin{cases}
			1& l=1 \\
			\mathop{\arg\max}\limits_{c}\sum_i\delta(\tilde{y}_{i}(\alpha_l(c)), y_i)& l\neq 1 .
		\end{cases}
		$$
		Here $\tilde{y}_{i}(\alpha_l(c))$ is the prediction of the ensemble classifier composed of the first $l$ classifiers. 
		$\mathcal{P}^{\text{true}}_{l}$ and $\mathcal{P}^{\text{false}}_{l}$ are defined in the main text.
		
		Update the weight of training set:
		\begin{equation}
			w_{l+1, i} = \frac{w_{l, i}}{{Z_{l+1}}}\,\text{exp}[P(x_i, \bm{\theta}^{*}_{l}) \cdot (1 - 2 \delta_{\Tilde{y}_{l, i} y_i})] \nonumber,
		\end{equation}
		with the normalization factor
		\begin{equation}
			{Z_{l+1}}=\sum_{i=1}^{N}w_{l, i}\,\text{exp}[P(x_i, \bm{\theta}^{*}_{l}) \cdot (1 - 2 \delta_{\Tilde{y}_{l, i} y_i})] \nonumber.
		\end{equation}
	}
\end{algorithm}

\vspace{.3cm}
\noindent\textbf{Training the QNN classifier}

\noindent We train the QNN classifier with epochs. At each epoch, we first shuffle the training set and then divide it evenly into 120 batches, with each batch containing 30 images. During the training procedure, we iterate through the 120 batches of dataset for each epoch, and train the QNN classifier for 7 epochs.
To encode an image sample $x_i$, we first flatten it into a $784$-dimensional vector $\bm{v}_i$, which is then divided by $255^2$ and transformed into a $30$-dimensional vector $\bm{x}_i$ by a trainable matrix $\bm{W}$.
The data vector $\bm{x}_i$ is encoded together with the trainable parameter $\bm{\theta}$ into the QNN circuit, as shown in Fig.~\ref{Fig2}b.
{We initialize $\bm{\theta}$ and $\bm{W}$ by randomly sampling each of their element from the Gaussian distribution $\mathcal{N}\left(\pi, \left(\frac{\pi}{3}\right)^{2}\right)$ and $\mathcal{N}\left(\frac{\pi}{60}, \left(\frac{\pi}{180}\right)^2\right)$, respectively}.
$\bm{W}$ and $\bm{\theta}$ are trained by using the gradient based Adam optimizer~\cite{kingma2017adammethodstochasticoptimization}, with the gradient experimentally measured by using the parameter shift rule~\cite{PhysRevA.98.032309}. The updating rules for $\bm{W}$ and $\bm{\theta}$ are detailed in Algorithm~\ref{alg:QNN}.

\begin{algorithm}
	\caption{Updating rules for $\bm{W}$ and $\bm{\theta}$.}
	\label{alg:QNN}
	\SetAlgoLined
	\KwIn{~\\
		Trainable matrix $\bm{W}_j$; \\
		Trainable parameter $\bm{\theta}_j$; \\
		Batch of training data $\left\{(\bm{v}_b, y_b)\right\}^{B}_{b=1}$; \\
		Batch of corresponding sample weight $\bm{w}=\{w_{b}\}^{B}_{b=1}$. \\
	}
	
	\KwOut{~\\
		Updated trainable matrix $\bm{W}_{j+1}$; \\
		Updated trainable parameter $\bm{\theta}_{j+1}$.
	}
	\hrulefill ~\\
	$\bm{g}_{\bm{\theta}_j}, \bm{g}_{\bm{W}_j} \leftarrow \bm{0}$~\\
	\For{$b = 1:B$}{
		Obtain the encoding data $\bm{x}_b =  \bm{W}_j \cdot \bm{v}_b\nonumber$.
		Measure the gradient 
		\begin{align}
			\nabla_{\bm{x}_b} \mathcal{L}_{b} &= \nabla_{\bm{\theta}_j} \mathcal{L}_{b} = \left[\frac{\partial \mathcal{L}_{b}}{\partial \theta_{j,i}}\right]_{i=1}^{N_t} \nonumber\\ &= -\frac{w_{b}}{P_{y_b}(\bm{x}_b, \bm{\theta}_{j})} \left[\frac{\partial P_{y_b}}{\partial \theta_{j,i}}\right]_{i=1}^{N_t} \nonumber\\&= -\frac{w_{b}}{P_{y_b}(\bm{x}_b, \bm{\theta}_{j})} \left[\frac{P_{y_b}(\bm{x}_b, \bm{\theta}^{+}_{j,i}) - P_{y_b}(\bm{x}_b, \bm{\theta}^{-}_{j,i})}{2}\right]_{i=1}^{N_t}\nonumber,
		\end{align}
		where $N_t$ is the number of trainable parameters and $\bm{\theta}^{\pm}_{j,i} = [\theta_{j,1}; \theta_{j,2}; ...; \theta_{j,i} \pm \pi/2; ...; \theta_{j,N_t}]$.
		
		Calculate the gradient of transform matrix:
		\begin{equation}
			\nabla_{\bm{W}_j} \mathcal{L}_{b}=  \nabla_{\bm{x}_b} \mathcal{L}_{b} \cdot \bm{v}_{b}^T\nonumber.
		\end{equation}
		
		$\bm{g}_{\bm{\theta}_j} = \bm{g}_{\bm{\theta}_j} + \nabla_{\bm{\theta}_j} \mathcal{L}_{b}$.
		
		$\bm{g}_{\bm{W}_j} = \bm{g}_{\bm{W}_j} + \nabla_{\bm{W}_j} \mathcal{L}_{b}$.
	}
	
	Calculate:
	\begin{equation}
		\bm{\theta}_{j+1} = \bm{\theta}_j + \mathcal{A}\left[\beta_1, \beta_2, \gamma, j, \bm{g}_{\bm{\theta}_j}\right] \nonumber,
	\end{equation}
	\begin{equation}
		\bm{W}_{j+1} = \bm{W}_j + \mathcal{A}\left[\beta_1, \beta_2, \gamma, j, \bm{g}_{\bm{W}_j}\right] \nonumber,
	\end{equation}
	where $\mathcal{A}$ is the Adam optimizer with $\beta_1 = 0.9$, $\beta_2 = 0.999$, and $\gamma = 0.5$.
	
\end{algorithm}

\vspace{.3cm}
\noindent\textbf{Training the QCNN classifier}

\noindent The training procedure of the QCNN classifier is similar to that of the QNN classifier. Each QCNN classifier is trained for $5$ epochs. At each epoch, the training set is shuffled and divided into $76$ batches, with each batch containing $29$ quantum states.
The quantum states can be directly input into the QCNN classifier.
The trainable parameter $\bm{\theta}$ is initialized by randomly generating each of its element in a range of {$[-\pi, \pi)$}.
The gradient of the loss function with respect to $\bm{\theta}$ is evaluated by using the finite difference method, with the updating rules detailed in Algorithm~\ref{alg:QCNN}.

\begin{algorithm}
	\caption{Updating rules for $\bm{\theta}$}
	\label{alg:QCNN}
	\SetAlgoLined
	\KwIn{~\\
		weights $\bm{w}=\{w_{b}\}^{B}_{b=1}$;
		Trainable parameter $\bm{\theta}_j$; \\
		Batch of training data $\left\{(x_b, y_b)\right\}^{B}_{b=1}$; \\
		Batch of corresponding sample weight $\bm{w}=\{w_{b}\}^{B}_{b=1}$.
	}
	
	\KwOut{~\\
		Updated trainable parameter $\bm{\theta}_{j+1}$.
	}
	\hrulefill ~\\
	
	$\bm{g}_{\bm{\theta}_j} \leftarrow \bm{0}$ ~\\
	\For{$b = 1:B$}{
		Measure the gradient 
		\begin{align}
			\nabla_{\bm{\theta}_j} \mathcal{L}_{b} &= \left[\frac{\partial \mathcal{L}_{b}}{\partial \theta_{j,i}}\right]_{i=1}^{N_t} \nonumber\\ &= -w_{b} \left[\frac{\partial \text{ln}(P_{y_b})}{\partial \theta_{j,i}}\right]_{i=1}^{N_t} \nonumber\\&= -w_{b} \left[\frac{\text{ln}(P_{y_b}(\bm{x}_b, \bm{\theta}^{+}_{j,i})) - \text{ln}(P_{y_b}(\bm{x}_b, \bm{\theta}^{-}_{j,i}))}{2\epsilon}\right]_{i=1}^{N_t}\nonumber,
		\end{align}
		where $N_t$ is the number of trainable parameters, $\bm{\theta}^{\pm}_{j,i} = [\theta_{j,1}; \theta_{j,2}; ...; \theta_{j,i} \pm \epsilon; ...; \theta_{j,N_t}]$, and $\epsilon = 0.2$.
		$\bm{g}_{\bm{\theta}_j} = \bm{g}_{\bm{\theta}_j} + \nabla_{\bm{\theta}_j} \mathcal{L}_{b}$.
	}
	Calculate:
	\begin{equation}
		\bm{\theta}_{j+1} = \bm{\theta}_j + \mathcal{A}\left[\beta_1, \beta_2, \gamma, j, \bm{g}_{\bm{\theta}_j}\right] \nonumber,
	\end{equation}
	where $\mathcal{A}$ is the Adam optimizer with $\beta_1 = 0.9$, $\beta_2 = 0.999$, and $\gamma = 0.02$ is the learning rate.
\end{algorithm}

\let\oldaddcontentsline\addcontentsline
\renewcommand{\addcontentsline}[3]{}
\bibliography{main_supp}
\let\addcontentsline\oldaddcontentsline

\resetlinenumber
\clearpage
\onecolumngrid

\begin{center} 

	{\large \bf Supplementary of ``Quantum ensemble learning with a programmable superconducting processor"}
\end{center} 
\maketitle

\newcounter{suppfigure}
\newcounter{suppalg}
\renewcommand{\thesuppfigure}{S\arabic{suppfigure}}
\renewcommand{\thesuppalg}{S\arabic{suppalg}}

\newcommand{\beginsupplement}{%
	\setcounter{table}{0}
	\renewcommand{\thetable}{S\arabic{table}}%

    \renewcommand{\thefigure}{\thesuppfigure}
    \setcounter{suppfigure}{0}
    
    \renewcommand{\thealgocf}{\thesuppalg}
    \setcounter{suppalg}{0}
    
	\setcounter{equation}{0}
	\renewcommand{\theequation}{S\arabic{equation}}%
	\setcounter{section}{0}
	\renewcommand{\thesection}{\arabic{section}}%
}
\setcounter{page}{1}
\beginsupplement
\renewcommand{\thepage}{S\arabic{page}}
\renewcommand{\bibnumfmt}[1]{[S#1]}
\tableofcontents

\section{Numerical simulation}

To demonstrate the efficiency of our Adaboost.Q algorithm, we compare it with the conventional Adaboost.M1 algorithm using numerical simulations.
The Adaboost.M1 algorithm is shown in Algorithm~\ref{alg:AdaBoostM1}.
The classification task considered here is a ten-class classification task, which is the same one as considered in the main text and carried out with the same structure of the QNN circuit.
Each ensemble classifier is composed of four base classifiers. We benchmark the accuracy of each algorithm for 500 times, with the initial trainable parameters randomly generated each time. The results are shown in Fig.~3d of the main text.

\begin{algorithm}
	\stepcounter{suppalg}
	\caption{AdaBoost.M1}
	\label{alg:AdaBoostM1}
	\SetAlgoLined
	\KwIn{ ~\\
		Training set $\mathcal{T}=\left\{(x_i, y_i)\right\}^{N}_{i=1}$, where $x_i \in \mathcal{X}$ and $y_i \in \mathcal{Y} \subseteq \left\{1, 2, ..., K\right\}$;\\
		The number of weak classifier $L$; \\
		A series of weak classifiers $\left\{G(x, \bm{\theta}_l)\right\}^{L}_{l=1}$,
		
		\begin{equation}
			G(x, \bm{\theta}_l): \mathcal{X} \rightarrow \bm{R}^K, \quad G(x_i, \bm{\theta}_l) = \left\{P_{k}(x_{i}, \bm{\theta}_l)\right\}_{k=1}^{K} \nonumber,
		\end{equation}
		with $P_{k}(x_i, \bm{\theta}_l)$ defined in the main text.
	}
	\KwOut{ ~\\
		Combined classifier
		\begin{equation}
			\mathcal{G}(x) = \sum_{l=1}^{L}\alpha_l \cdot G(x, \bm{\theta}^*_l) \nonumber,
		\end{equation}
		where $\alpha_{l}$ and $\bm{\theta}^*_l$ are the weight and optimal parameters of the $l$th classifier.
		The output label of the combined classifier is
		\begin{equation}
			\tilde{y}_i = \mathop{\arg\max}_{k}\mathcal{G}(x_i) \nonumber.
		\end{equation} 
	}
	\hrulefill ~\\
	
	Initialize the weights of the training set,
	\begin{equation}
		\bm{w}_1 = \left\{w_{1,i}\right\}_{i=1}^{N}, \nonumber
	\end{equation}
	where $w_{1, i} = 1 / N$.
	
	\For{$l = 1:L$}{
		Train the classifier $G(x, \bm{\theta})$ with the weighted training set.
		Obtain
		\begin{equation}
			\tilde{y}_{l, i} = \mathop{\arg\max}\limits_{k}G(x_i, \bm{\theta}^{*}_{l}) \nonumber,
		\end{equation}
		with error rate:
		\begin{equation}
			e_l = \sum^{N}_{i=1}w_{l,i}\,\delta(\tilde{y}_{l,i}, y_i) \nonumber.
		\end{equation}
		
		Calculate the $l$th classifier's weight:
		\begin{equation}
			\alpha_l = \text{ln}\left(\frac{1 - e_l}{e_l}\right) + \text{ln}(K - 1). \nonumber
		\end{equation}
		
		Update the weight of training set:
		\begin{equation}
			w_{l+1, i} = \frac{w_{l, i}}{{Z_{l+1}}}\,\text{exp}(\alpha_l \delta_{\Tilde{y}_{l, i} y_i}) \nonumber,
		\end{equation}
		with the normalization factor
		\begin{equation}
			{Z_{l+1}}=\sum_{i=1}^{N}w_{l, i}\,\text{exp}(\alpha_l\delta_{\Tilde{y}_{l, i} y_i}) \nonumber.
		\end{equation}
	}
\end{algorithm}

\section{Experimental information}

\subsection{Device information}
\begin{figure*}[h]
\stepcounter{suppfigure}
    \includegraphics[width=1.0\textwidth]{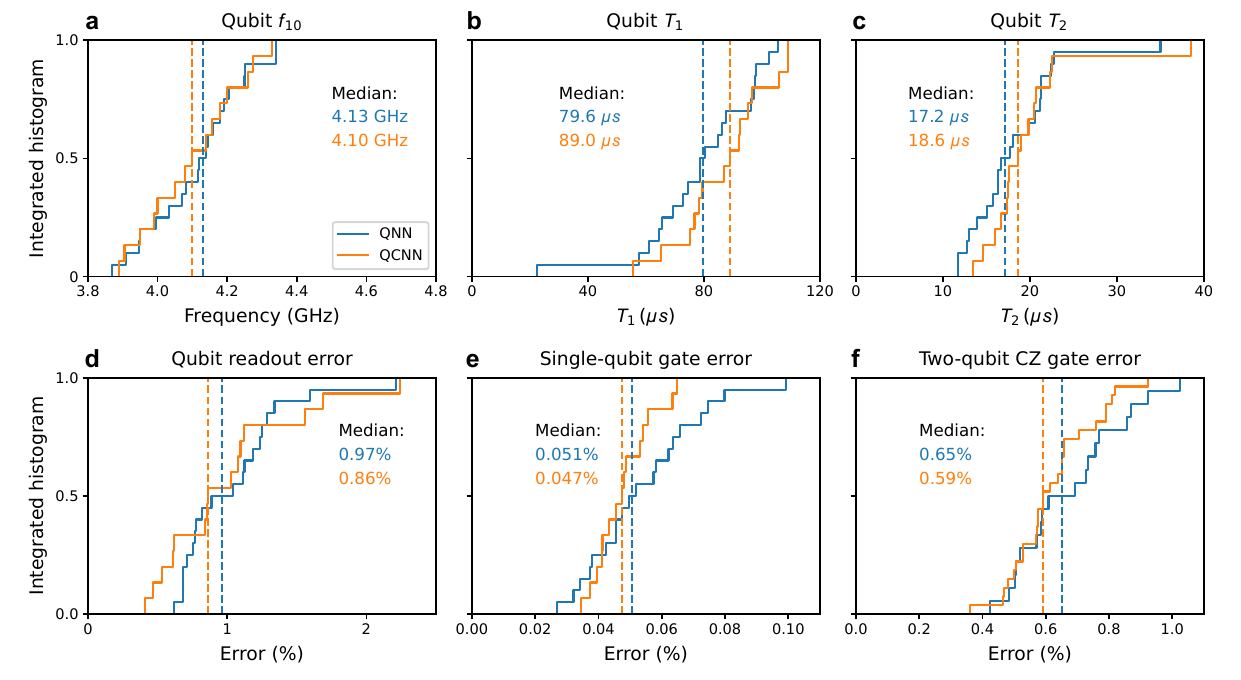}
    \caption{\label {supp:fig_device_params}
    {\bf Integrated histograms of qubit parameters.}
 \textbf{a}, Qubit idle frequency.
 \textbf{b}, Qubit relaxation time measured at the idle frequency.
 \textbf{c}, Qubit dephasing time measured using Hahn echo sequence.
 \textbf{d}, Readout fidelity of the qubit.
 \textbf{e}, Pauli error of simultaneous single-qubit gate.
 \textbf{f}, Pauli error of two-qubit CZ gate.
 Dashed lines indicate the median values.
}
\end{figure*}

Our experiments are performed on a superconducting quantum processor, which has $11 \times 11$ transmon qubits aligned in a  square grid, with the nearest neighboring qubits connected by tunable couplers~\cite{xu2023digital}.
Each qubit has individual microwave control for realizing single-qubit gates and flux control for frequency modulation. Each coupler has an individual flux control for tuning the coupling strength between the corresponding two neighboring qubits.
As illustrated in Fig.~2 and Fig.~4 of the main text, we use two 10-qubit chains and a 15-qubit chain for realizing the quantum neural network (QNN) and the quantum convolutional neural network (QCNN) in our experiment, with their properties and gate errors shown in Fig.~\ref{supp:fig_device_params}.  

\subsection{Quantum data preparation}
For the quantum phase recognition task in the main text, we use variational quantum circuits (VQCs) to prepare the approximate ground states in the training and test sets.
A general structure of the VQC is composed of blocks of single- and two-qubit gates.
Each block contains six layers of parameterized single-qubit gates and two layers of two-qubit CZ gates, as shown in Fig.~\ref{supp:fig_state_preparation}.
We use three and four blocks to prepare the quantum states in the training and test sets, respectively. The circuit is implemented on the quantum processor before applying the QCNN circuit.

For a given model parameter set $\{h_1, h_2\}$, we determine the parameters in the VQC by minimizing the energy expectation value of the corresponding Hamiltonian in a classical computer by using the gradient-based Nadam optimizer.
The variational parameters are initialized to be the optimized ones of the neighboring point if there existed one, otherwise with each element randomly generated in the range of {$[-\pi, \pi)$}. The optimization procedure ends when the energy value no longer decreases in the next 20 iteration steps. 
The circuit is accepted when 1). the energy differences between the neighboring steps is less than $5\times 10^{-5}$ for 20 iterations; 2). the difference between the optimized energy and the ideal ground state energy obtained by the DMRG (Density Matrix Renormalization Group) method is less than 0.5; and 3). the corresponding order parameters of the optimized states in each of the three phases meet the following criteria: a). $\langle S\rangle>0.2$ for states in SPT phase; b). $\langle S\rangle<0.2$ and $\langle X_7X_8\rangle>0$ for states in FM phase; and c).  $\langle S\rangle<0.2$ and $\langle X_7X_8\rangle<0$ for states in Ising phase.
If not accepted, the optimization will be retried with newly and randomly generated variational parameters.
The model parameter sets that fail 10 retries will be discarded in this work, which usually locate around the critical points.
The order parameters of the numerically obtained quantum states in the training and test sets are shown in Fig.~\ref{supp:fig_s_xx}.
As a result, we obtain approximate ground states with their energies locating within the range of $1.8\%$ above the ground state energies, as shown in Fig.~\ref{supp:energy}.

\subsection{More experimental data}
\begin{figure*}[!t]
\stepcounter{suppfigure}
\centering
\includegraphics[width=1.0\textwidth]{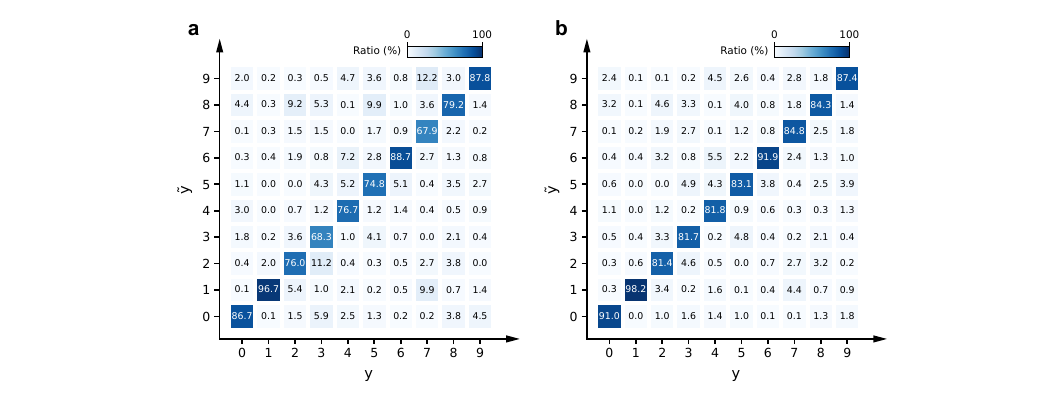}
\caption{
\label {supp:fig_classify_compare}
{\bf Comparison of detailed classification results between single classifier and ensemble classifier.}
The classification performance of the trained model on the whole MNIST test set, where the ratios of classifying $y$ to $\tilde{y}$ are displayed.
\textbf{a}, Single classifier's classification result.
\textbf{b}, Ensemble classifier's classification result.
}
\end{figure*}

\begin{figure*}[h]
\stepcounter{suppfigure}
\centering
\includegraphics[width=1.0\textwidth]{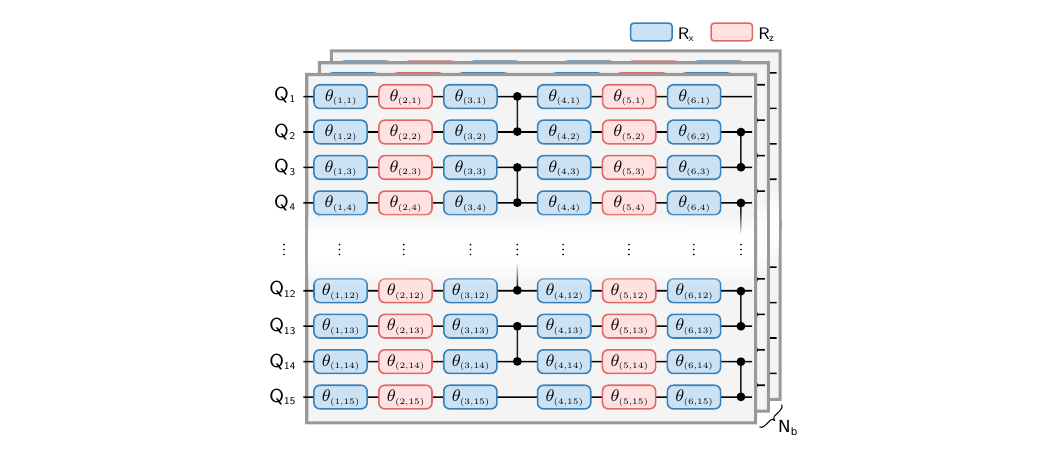}
\caption{
\label {supp:fig_state_preparation}
{\bf Varartional quantum circuit for preparing the quantum phase.}
The circuit consists of $\textsf{N}_\textsf{b}$ blocks, each of which is composed of six layers of single-qubit gates inserted by two layers of two-qubit CZ gates.
The single-qubit gate rotation angles serve as variational parameters.
The quantum states in the training and test sets are generated with block numbers $\textsf{N}_\textsf{b}=3$ and $4$, respectively.
} 
\end{figure*}

\begin{figure*}[h]
\stepcounter{suppfigure}
\centering
\includegraphics[width=1.0\textwidth]{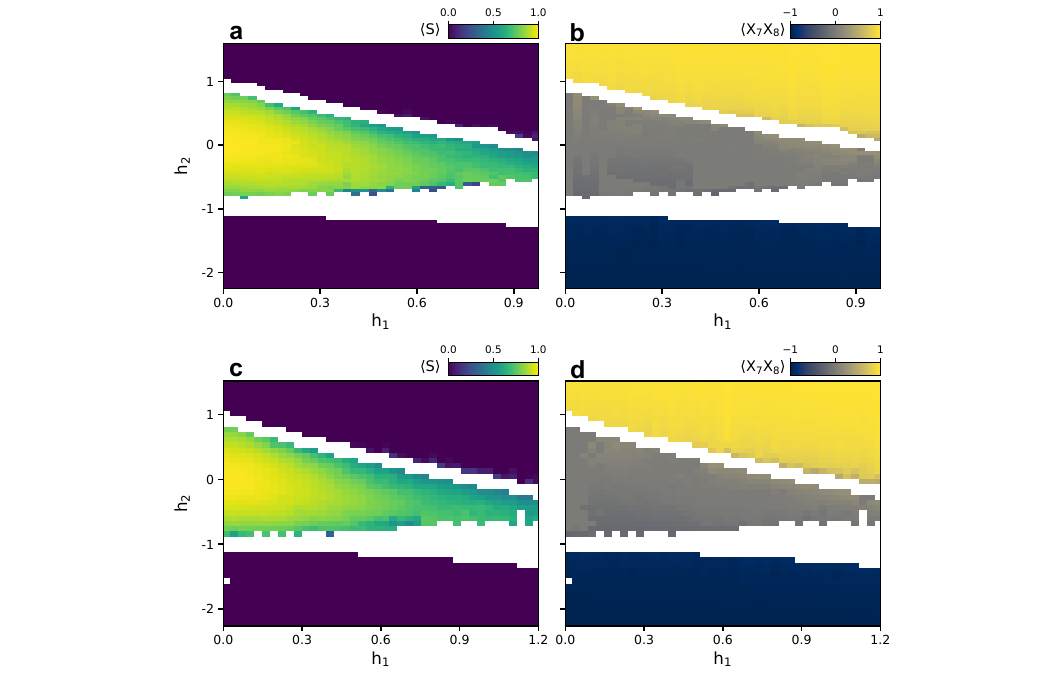}
\caption{
\label {supp:fig_s_xx}
{\bf $\langle S \rangle$ and $\langle X_{7}X_{8} \rangle$ of quantum states generated by VQCs during the numerical simulation.}
Expectation values of the observables $S$ and $X_7X_8$ of the quantum states in training and test set are shown in (a), (b) and (c), (d), respectively.
} 
\end{figure*}

\begin{figure*}[h]
\stepcounter{suppfigure}
\centering
\includegraphics[width=1.0\textwidth]{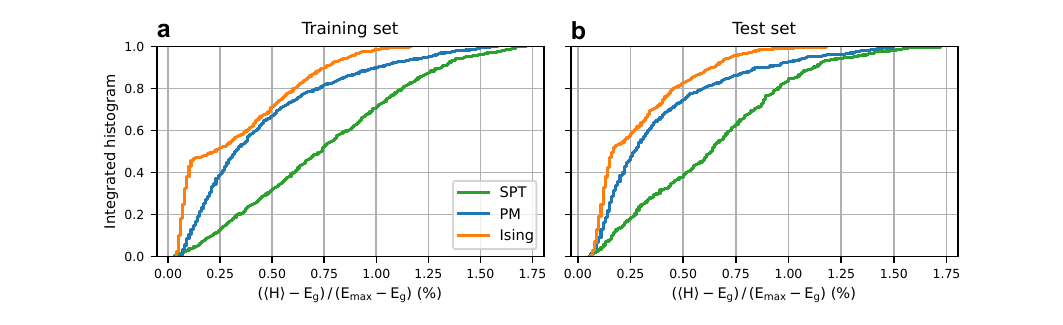}
\caption{
\label {supp:energy}
{\bf Integrated histograms of the energy deviation from the ground states for the variationally generated quantum states in the three phases.} Here $\langle H\rangle$, $E_\text{max}$, and $E_\text{g}$ represent the energy of the approximate ground states, the maximum eigen energy and the energy of the ideal ground states, respectively.
} 
\end{figure*}

The detailed classification results for single and ensemble classifiers on the MNIST hand-written digits dataset in the main text are displayed in Fig~\ref{supp:fig_classify_compare}.

\end{document}